# The Diversity-Innovation Paradox in Science

Bas Hofstra[1]*, Vivek V. Kulkarni[2], Sebastian Munoz-Najar Galvez[1], Bryan He[2], Dan Jurafsky[2], & Daniel A. McFarland[1]*

[1] *Stanford University, Graduate School of Education,*
*520 Galvez Mall, Stanford, CA 94305, USA*
[2] *Stanford University, Department of Computer Science,*
*353 Serra Mall, Stanford, CA 94305 USA*

* To whom correspondence should be addressed: bhofstra@stanford.edu or dmcfarla@stanford.edu

**Classification**

Social Sciences

**Significance Statement**

By analyzing data from nearly all US PhD-recipients and their dissertations across three decades, this paper finds demographically underrepresented students innovate at higher rates than majority students, but their novel contributions are discounted and less likely to earn them academic positions. The discounting of minorities' innovations may partly explain their underrepresentation in influential positions of academia.

**Abstract**

Prior work finds a diversity paradox: diversity breeds innovation, and yet, underrepresented groups that diversify organizations have less successful careers within them. Does the diversity paradox hold for scientists as well? We study this by utilizing a near-population of ~1.2 million US doctoral recipients from 1977-2015 and following their careers into publishing and faculty positions. We use text analysis and machine learning to answer a series of questions: How do we detect scientific innovations? Are underrepresented groups more likely to generate scientific innovations? And are the innovations of underrepresented groups adopted and rewarded? Our analyses show that underrepresented groups produce higher rates of scientific novelty. However, their novel contributions are devalued and discounted: e.g., novel contributions by gender and racial minorities are taken up by other scholars at lower rates than novel contributions by gender and racial majorities, and equally impactful contributions of gender and racial minorities are less likely to result in successful scientific careers than for majority groups. These results suggest there may be unwarranted reproduction of stratification in academic careers that discounts diversity's role in innovation and partly explains the underrepresentation of some groups in academia.

**Introduction**

Innovation drives scientific progress. Innovation propels science into uncharted territories and expands humanity's understanding of the natural and social world. Innovation is also believed to be predictive of successful scientific careers: innovators are science's trailblazers and discoverers, so producing innovative science may lead to successful academic careers (*1*). At the same time, a common hypothesis is that demographic diversity brings such innovation (*2-5*). Scholars from underrepresented groups have origins, concerns, and experiences that differ from groups traditionally represented, and their inclusion in academe diversifies scholarly perspectives. In fact, historically underrepresented groups often draw relations between ideas and concepts that have been traditionally missed or ignored (*4-7*). Given this, if demographic groups are unequally represented in academia, then one would expect underrepresented groups to generate more scientific innovation than overrepresented groups and have more successful careers (see Supplementary Text). Unfortunately, the combination of these two relationships – diversity-innovation and innovation-careers – fails to result and poses a paradox. If gender and racially underrepresented scholars are likely to innovate *and* innovation supposedly leads to successful academic careers, then how do we explain persistent inequalities in scientific careers between minority and majority groups (*8-13*)? One explanation is that the scientific innovations produced by some groups are discounted, possibly leading to differences in scientific impact and successful careers.

In this paper, we set out to identify the diversity-innovation paradox in science and explain why it arises. We provide a system-level account of science using a near-complete population of US doctorate recipients (~1.2 million) where we identify scientific innovations (*14-19*) and analyze the rates at which different demographic groups relate scientific concepts in novel ways, the extent to which those novel conceptual relations get taken up by other scholars, how "distal" those linkages are (*14*), and the subsequent returns they have to scientific careers. Our analyses use observations spanning three decades, all scientific disciplines, and all US doctorate awarding institutions. Through them we are able to (a) compare minority scholars' rates of scientific novelty vis-à-vis majority scholars, and then ascertain whether and why their novel conceptualizations (b) are taken up by others, and in turn, (c) facilitate a successful research career.

**Innovation as Novelty and Impactful Novelty in Text**

Our dataset stems from ProQuest dissertations (*20*), which includes records of nearly all US PhD theses and their metadata from 1977-2015: student names, advisors, institutions, thesis titles, abstracts, disciplines, etc. These structural and semantic footprints enable us to consider students' rates of innovation at the very onset of their scholarly careers and their academic trajectory afterwards – i.e., their earliest conceptual innovations and how they correspond with successful academic careers (*21*). We link these data with several data sources to arrive at a near-ecology of US PhD students and their career trajectories. Specifically, we link ProQuest dissertations to the US Census data (2000 and 2010) and Social Security Administration data (1900-2016) to infer demographic information on students' gender and race (i.e., name signals for white, Asian, or underrepresented minority [Hispanic, African American, or Native American], see Materials and Methods and Supplementary Text); we link ProQuest dissertations to Web of Science – a large-scale publication database with ~38 million academic publications (1900-2017) – to find out which students have continued research careers; and we weigh our inferential analyses by population records of the number of PhD recipients for each distinct university-year combination to render results generalizable to the population (see Supplementary Text).

To measure scientific innovation, we first identify the set of scientific concepts being employed in theses. For this, we use natural language processing techniques of phrase extraction and structural topic modeling (*22, 23*) to identify terms representing substantive concepts in millions of documents (*# concepts*, Mean = 56.500; Median = 57; SD = 19.440, see Materials and Methods and Table S1) (*24*). Next, we filter and identify when pairs of meaningful concepts are first related to one another in a thesis. By summing the number of novel conceptual co-occurrences within each thesis, we develop a measure of how conceptually novel a thesis and author is (*# new links*) – their *novelty*. However, not all novel conceptual linkages are taken up in ensuing works and have the same impact on scholarship. To capture *impactful novelty* we measure how often a thesis' new conceptual linkages are adopted in ensuing documents of each year (*uptake per new link*) (see Figure 1).

Our broad perspective on innovation mirrors key theoretical perspectives on scientific innovation, where "science is the constellation of facts, theories, and methods collected in current texts (*25*)." Scientific development is then the process where concepts are added to the ever-growing "constellation" – i.e., our accumulating corpus of texts – in new combinations, or the introduction of new links between scientific concepts (*14, 15, 25-27*). As such, the combination of novelty as the number of unique recombinations of scientific concepts (*# new links*, Mean = 9.026; Median = 4; SD = 13.744; 20.9% of students do not introduce links) and impactful novelty as the average future adoption of these unique recombinations (*uptake per new link*, Mean = .790; Median = .333, SD = 3.079) reflect different notions of scientific innovation. Novelty in itself does not automatically imply innovation, nor is the future adoption of novelty a prerequisite to innovation – e.g., *which* novelty gets adopted may be in itself a function of structural processes. The advantage of our focus on conceptual recombination compared to citation metrics for innovation is that it is insensitive to (a) prioritizing some academic disciplines over others with regard to journal indexing and (b) the plethora of reasons as to why scholars cite other work (*28, 29*).

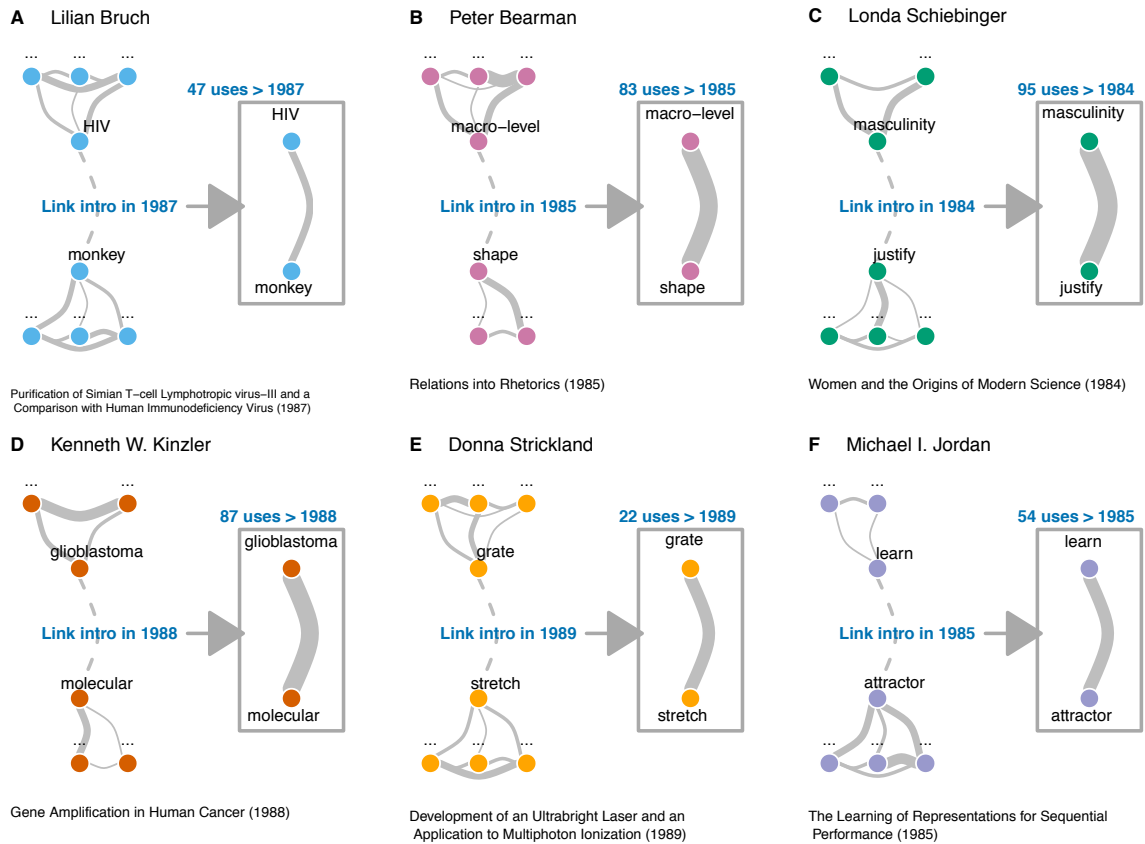

**Figure 1.** *The introduction of innovations and their subsequent uptake.*
*(A-F) Examples drawn from the data illustrate our measures of novelty and impactful novelty. Nodes represent concepts and edge thickness indicates the frequency of their co-usage. Students can introduce new links (dotted lines) as their work enters the corpus. These examples concern novel links taken up at significantly higher rates than usual (e.g., 95 uses of Schiebinger's link after 1984). The mean (median) uptake of new links is .790 (.333), and ~50% of new links never gets taken up. (A) Lilian Bruch was among the pioneering HIV researchers (30) and her thesis introduced the link between "HIV" and "monkeys", indicating innovation in scientific writing as HIV's origins are often attributed to non-human primates. (C) Londa Schiebinger was the first to link "masculinity" with "justify," reflecting her pioneering work on gender bias in academia (31). (E) Donna Strickland won the 2018 Nobel Prize in Physics for her PhD work on chirped pulse amplification, utilizing grating-based stretchers and compressors (32).*

**Results**

Who introduces novelty and whose novelty is impactful? We first model individual rates of novelty (# new links) and impactful novelty (uptake per new link) by several notions of demographic diversity: the gender and racial representation in a students' discipline and by gender/race indicators reflecting historically underrepresented groups (see Figure 2). We keep institution, academic discipline, and graduation year constant (*33, 34*) (see Materials and Methods, Supplementary Text, Figures S1 and S4, and Table S2). We find that the more students are underrepresented genders ($p < .001$) or races ($p < .05$) in their discipline, the more they are likely to introduce novel conceptual linkages (# new links). Yet, the more students are surrounded by peers of a *similar* gender in their discipline, the more their novel conceptual linkages are taken up by others ($p < .01$) – i.e., the less a students' gender is represented, the less their novel contributions are adopted by others (uptake per new link). Findings for binary gender and race indicators follow similar patterns. Women and non-white scholars introduce more novelty (both $p < .001$) but have less impactful novelty (both $p < .05$) when compared to men and white students. Additionally, intersectional analyses of gender-race combinations suggest that non-white women, white women, and non-white men all have higher rates of novelty compared to white men (all $p < .001$), but that white men have higher levels of impactful novelty compared to the other groups (all $p < .01$). Combined, these findings suggest that demographic diversity breeds novelty, and especially historically underrepresented groups in science introduce novel recombinations, but their rate of adoption by others is lower, suggesting their novel contributions are discounted.

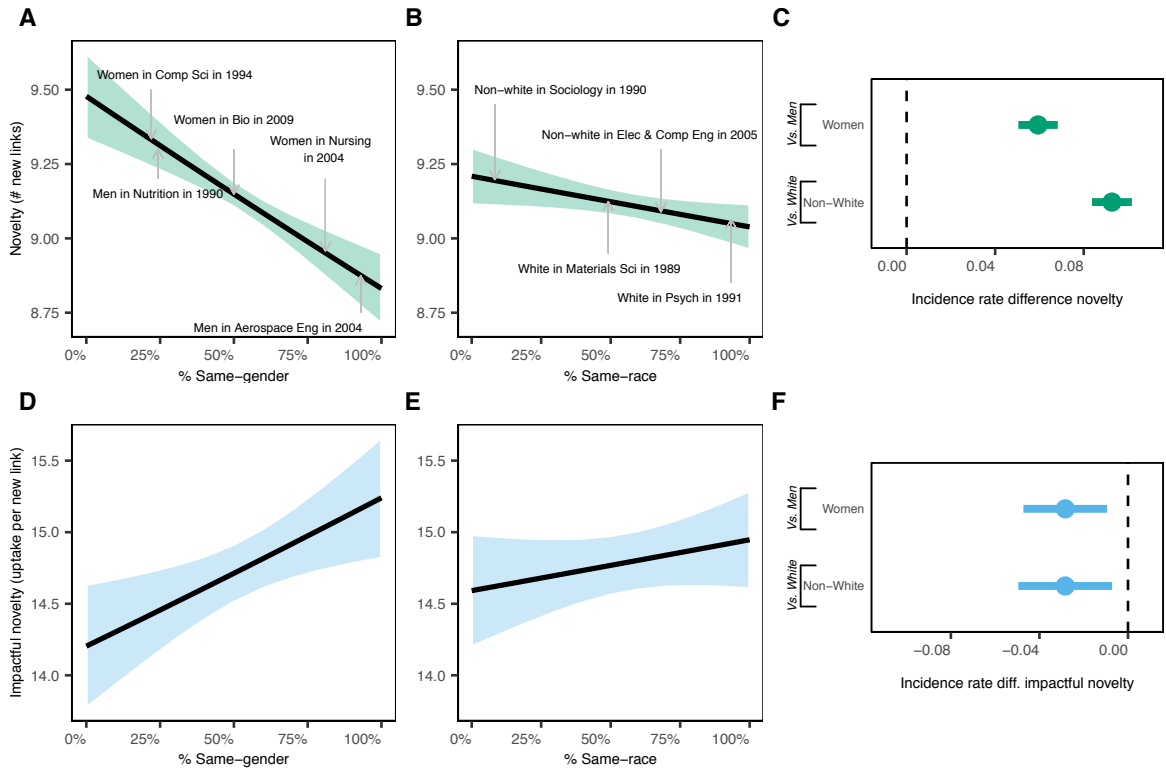

**Figure 2.** *Gender and race representation relate to novelty and impactful novelty.*
*(A) Introduction of novelty (# new links) by the percentage of peers with a similar gender in a discipline (N = 808,375). Specifically, the results suggest that the more students' own gender is underrepresented, the more novelty they introduce. (B) Similarly, the more students' own race is underrepresented, the more novelty they introduce. (C) Binary gender and race indicators suggest that historically underrepresented groups in science (women, non-white scholars) introduce more novelty (i.e., their incidence rate is higher). (D) In contrast, impactful novelty decreases as students have fewer peers of a similar gender and suggests underrepresented genders have their novel contributions discounted (N = 345,257). (E) There is no clear relation between racial representation in a discipline and impactful novelty. (F) Yet, the novel contributions of women and non-white scholars are taken up less by others than those of men and white students (their incidence rate is lower).*

So why is the novelty introduced by (historically) underrepresented groups less impactful? We test the common hypothesis that innovations that draw together concepts from very different fields or using distal metaphorical links receive less reward. We first identify how semantically *distal* or *proximal* newly linked concepts are from one another in the space of accumulated concepts using word embedding techniques (*35*) (see Figure 3, detailed in Materials and Methods). Word embedding techniques enable us to estimate the semantic location of concepts in a vast network of interrelated concepts and compare how distally (or proximally) positioned newly linked concepts are to one another in that space using cosine distance. For the set of newly linked concepts in each thesis, we average their semantic distance, and model whether some groups introduce more distal forms of novelty in their theses than other groups. We find that students whose gender is underrepresented in a discipline introduce slightly more concept linkages that are semantically distant (see Figure 3C; $p < .001$) and women introduce more distal novelty in comparison to men ($p < .001$). In turn, distal novelty relates inversely to impactful novelty; more distal new links between concepts receive far less uptake (see Figure 3D; $p < .001$). Hence, underrepresented groups introduce novelty, and the discounting of their novel contributions may be partly explained by how distal the conceptual linkages are that they introduce.

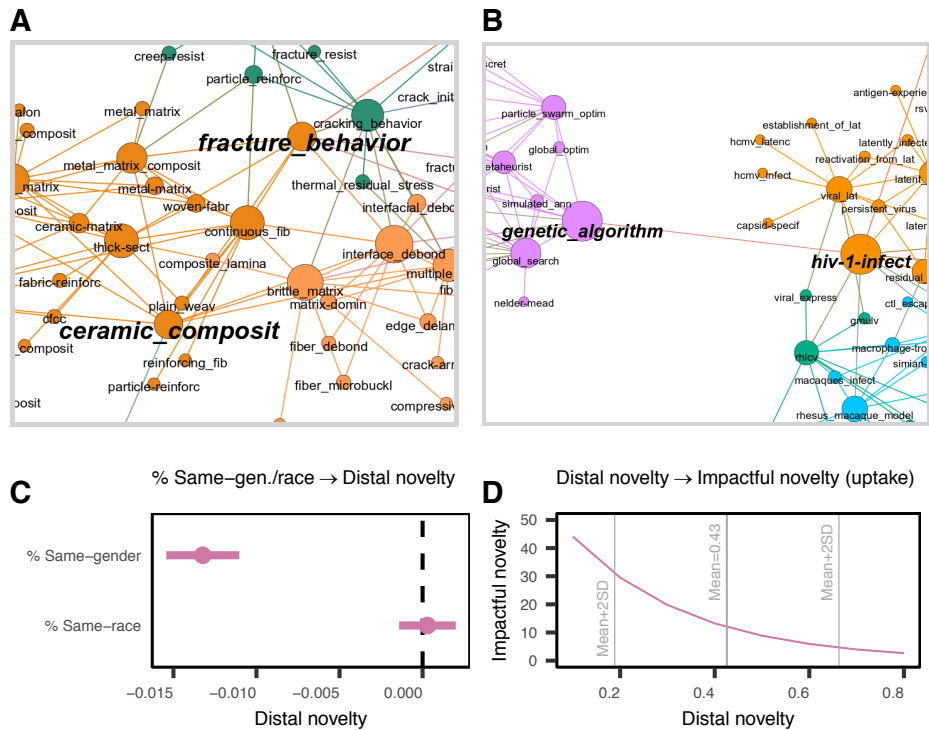

**Figure 3.** *Underrepresented genders introduce distal novelty, and distal novelty has less impact. (A-B) Communities (colors) of concepts and their linkages. (A) The link between "fracture_behavior" and "ceramic_composition" arises within a semantic cluster. Both concepts*

*are proximal in the embedding space of scientific concepts, and as such, their distal novelty score is low. (B) In contrast, the conceptual link between "genetic_algorithm" and "hiv-1" spans distinct clusters in the semantic network. As such, the concepts are distal in the embedding space of scientific concepts, and their distal novelty score is high. (C) Students of an overrepresented gender introduce more proximal novelty, and students from an underrepresented gender introduce more distal novelty in their theses. (D) In turn, the average distance of new links introduced in a thesis is negatively related to their future uptake.*

Finally, we examine how levels of novelty and impactful novelty relate to extended faculty and research careers. We model careers as (a) obtaining a research faculty position, and (b) as continuing research endeavors (see Figure 4 and Table S2). The former reflects PhDs who go on to become primary faculty advisors of PhDs at US research universities, while the latter reflects the broader pool of PhDs who continue to conduct research even if they do not have research advisor roles (e.g., in industry, non-tenure line role, etc.). For the latter, we identify which students become publishing authors in the Web of Science (*36*) five years after obtaining their PhD. The conceptual novelty and impactful novelty of a student's thesis is positively related to their likelihood of becoming both a research faculty member or continued researcher (all $p < .001$). This suggests that students are more likely to become professors and researchers if they introduce novelty or have impactful novelty.

However, consistent with prior work (*8-13*), we find that gender and racial inequality in scientific careers persists even if we keep novelty and impactful novelty constant (as well as year, institution, and discipline). Numerically underrepresented genders have lower odds of becoming research faculty (approximately 5% lower odds) and sustaining research careers (6% lower odds) compared to gender majorities (all $p < .001$). Similarly, numerically underrepresented races have lower odds of becoming research faculty (25% lower odds) and continuing research endeavors (10% lower odds) compared to majorities (all $p < .001$). Most surprisingly, the positive correlation of novelty and impactful novelty on career recognition varies by gender and racial groups and suggests underrepresented groups' innovations are discounted. The long-term career returns for novelty and impactful novelty are often lower for underrepresented rather than overrepresented groups. At a low level of (impactful) novelty gender minorities and majorities have approximately similar probabilities of faculty careers. But with increasing (impactful) novelty the probabilities diverge at the expense of gender minorities' chances (both slope differences $p < .01$). For instance, a 2SD increase from the median of (impactful) novelty increases the difference in probability of becoming a faculty researcher for gender minorities and majorities from about 3.5% (4.3%) to 9.5% (15%). These results hold over and above of the distance between newly linked concepts. This innovation discount also holds for traditionally underrepresented groups (i.e., women versus men, non-white versus white scholars).

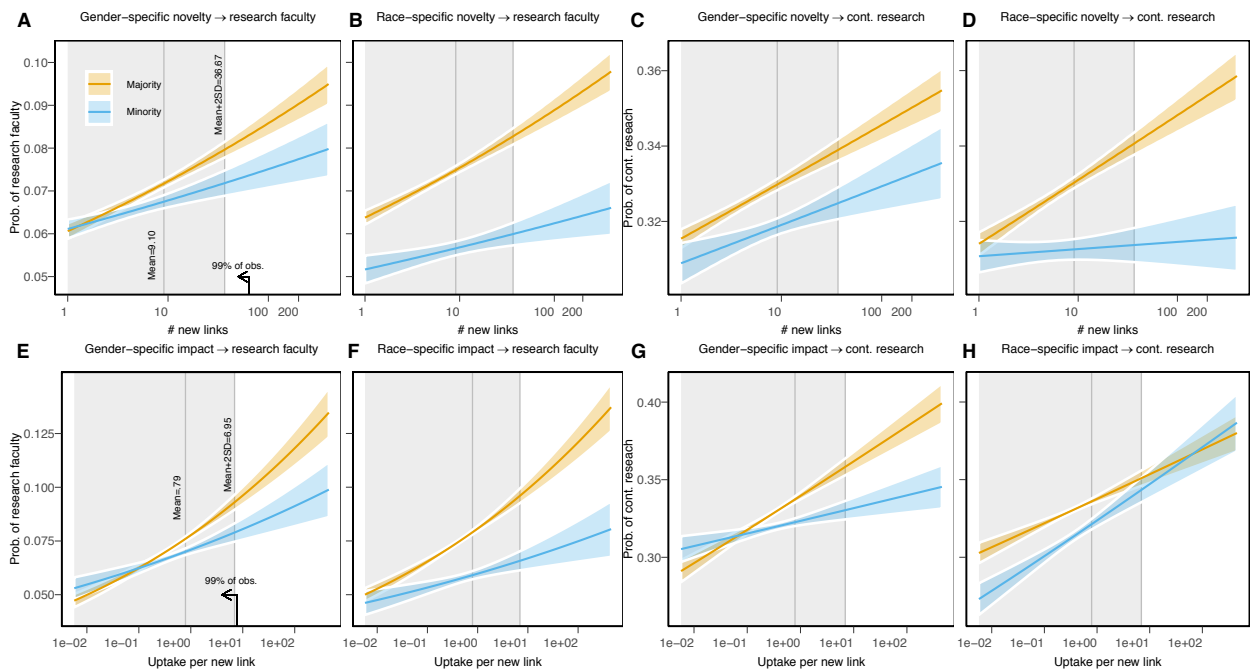

**Figure 4.** *The novelty and impactful novelty minorities introduce has discounted returns for their careers.*
(A-D) *Correlation of gender- and race-specific novelty with becoming research faculty or continued researcher (N = 805,236). As novelty increases, the probabilities of becoming faculty (for gender and race) and continuing research (for race) has diminished returns for minorities.*

*For instance, a 2SD increase from the median level of novelty (# new links) increases the difference in probability to become research faculty between gender minorities and majorities from 3.5% to 9.5%. (E-F) Correlation of gender- and race-specific impactful novelty with becoming research faculty and a continued researcher (when novelty is nonzero, N = 628,738). With increasing impactful novelty, the probabilities of becoming faculty (for gender and race) and continuing research (for gender) start to diverge at the expense of the career chances of minorities. For instance, a 2SD increase from the median of impactful novelty (uptake per new link) increases the difference in probability of becoming research faculty between gender minorities and majorities from 4.3% to 15%.*

**Discussion**

In this paper, we identified the diversity-innovation paradox in science. Consistent with intuitions that diversity breeds innovation, we find higher rates of novelty across several notions of demographic diversity (*2-7*). However, novel conceptual linkages are not uniformly adopted by others. Their adoption depends on which group introduces the novelty. For example, underrepresented genders have their novel conceptual linkages discounted and receive less uptake than the novel linkages presented by the dominant gender. Traditionally underrepresented groups in particular – women and non-white scholars – find their novel contributions receive less uptake. For gender minorities, this is partly explained by how "distal" the novel conceptual linkages are that they introduce. Entering science from a new vantage may generate distal novel connections that are difficult to integrate into localized conversations within prevailing fields. Moreover, this discounting extends to minority scientific careers. While novelty and impactful novelty both correspond with successful scientific careers, they offer lesser returns to the careers of gender and racial minorities than their majority counterparts (*8-13*). Specifically, at low impactful novelty we find that minorities and majorities are often rewarded similarly, but even highly impactful novelty is increasingly discounted in careers for minorities compared to majorities. And this discounting holds over and above of how distal minorities' novel contributions are.

In sum, this article provides a system-level account of innovation, and how those innovations differentially affect the scientific careers of demographic groups. This account is given for all academic fields from 1982-2010 by following over a million US students' careers and their earliest intellectual footprints. We reveal a stratified system where underrepresented groups have to innovate at higher levels to have similar levels of career success. These results suggest that the science careers of underrepresented groups end prematurely despite their crucial role in generating novel conceptual discoveries and innovation. Which trailblazers has science missed out on as a consequence? This question stresses the continued importance of critically evaluating and addressing biases in faculty hiring, research evaluation, and publication practices.

**Materials and Methods**

*Data*

This study focuses on a dataset of ProQuest dissertations filed by U.S. doctorate-awarding universities from 1977 to 2015 (*20*). The dataset contains 1,208,246 dissertations and accompanying dissertation metadata such as the name of the doctoral candidate, year awarded, university, thesis abstract, primary advisor (37.6% of distinct advisors mentor one student), etc. These data cover approximately 86% of all awarded doctorates in the US over three decades across all disciplines. We describe below how we follow PhD recipients going on into subsequent academic and research careers.

*Concept extraction from scientific text*

How do we extract concepts from text? Not all terms are scientifically meaningful; combining function words like "thus," "therefore," and "then", is substantively different from combining terms from the vocabulary of a specific research topic, like "HIV" and "monkey." We argue that innovation entails combining relevant terms from topical lexicons. Hence, we set out to define the latent themes in our corpus of dissertations and the most meaningful concepts in every theme. We employ Structural Topic Models (STMs) (*22*), commonly used to detect latent thematic dimensions in large corpora of texts (see Supplementary Text).

We fit topic models at K = [50-1000]. Fit metrics (see Figure S1 and Supplementary Text) plateau at K = 400, 500, and 600, and we use those three in this paper. To extract concepts, we identify the terms of relative importance to each latent theme in the dissertation corpus. Using the STM output, we obtain terms that are most-frequent and most-exclusive within a topic. This helps identify concepts that are both common and distinctive to balance generality and exclusivity. To get at this, we extract the top terms based on their FREX-score (*24*). FREX-scores compound the weighted frequency and exclusivity of a term in a topic. Here we explore three weighing schemes: equally balancing frequency and exclusivity (50/50), attaching more weight to frequency and less to exclusivity (75/25), and attaching more weight to exclusivity and less to frequency (25/75). As such, we analyze nine hyperparameter scenarios (three K and three FREX scenarios) for which sensitivity analyses provide robust results (see Table S2). For the results depicted in the main text, we report the scenario where frequency and exclusivity are equally balanced at K = 500.

We use all doctoral abstracts (1977-2015) as input documents for a semantic signal for the students' scholarship at the onset of their careers. However, in our inferential analyses, we utilize theses from 1982 to 2010 to a) allow for the scientific concept space to accumulate five years before we measure which students start to introduce links and b) to allow for the most recently graduated students (up until 2010) to have opportunities (five years) for their novelty to be taken up. Additionally, Figure S2 suggests that the "stable" years for link introductions and uptake per new link start at approximately 1982. The year fixed-effects in our inferential analyses (detailed below) further account for left- and right-censoring – i.e., year fixed-effects enable comparisons of students within rather than across years. Figure S3 depicts four exemplary topics and their concepts resulting from the structural topic models.

*Outcome variable–Novelty and impactful novelty*

Using the extracted scientific concepts, we aggregate co-occurring concepts in abstracts for each year, identifying which students first introduce each novel link. We remove spurious links (due to chance, combinations of extremely rare terms, etc.) by computing a significance score for each link: the log-odds ratio of the probability of link-occurrence (computed over all extracted concepts and all documents in the corpus) to the probability of each component concept term occurring independently over the corpus (*37*, detailed in the Supplementary Text). In sum, we identify "meaningful" links by filtering the documents for the top FREX terms via structural topic models, and then filtering for spurious links through a link significance score. If a link is introduced by two students in the same year they both get counted. (The percentage of links concurrently introduced is only 1.6% and the majority of concurrent link introductions arise from students getting their doctorate in the same year (99.7%).) This metric – the number of new link introductions – we call the *novelty* of a student's thesis (*# new links*, Mean = 9.026; Median = 4; SD = 13.744; 20.9% of students do not introduce new links).

Second, we measure *impactful novelty,* the uptake of a thesis' new links in ensuing theses. We count the total number of times theses in *following* years use the links first introduced by a prior thesis, normalized by the number of new links. We use the resulting metric, *uptake per new link* (Mean = .790; Median = .333, SD = 3.079), to quantify the average scientific impact for an individual student's thesis. See Figure S2 for the distributions and correlations of these outcome variables across the different K and FREX scenarios. Both metrics positively correlate with publication productivity and citation among those students that publish (see Table S3).

*Outcome variable–Distal novelty*

Some links are "distal" in that they link concepts that are located in distinct clusters of co-occurring concepts. Other links are "proximal" because they link concepts in the same semantic cluster or proximate location. For instance, *genetic_algorithm–hiv-1* is distal because it links concepts from distinct research areas: "genetic algorithms" (evolutionary computing) with "hiv-1" (medicine). In contrast, *fracture_behavior–ceramic_composition* is proximal because they are concepts from the same field.

To operationalize this notion of semantic distance, we embed each concept in a semantic network of cumulated co-occurring concepts, and then estimate its location in a vector space, representing each concept "$c$" by a fixed dimension vector (or "embedding") $v(c)$. We use the Skip-gram model (*35*), a standard approach that models co-occurrences between concepts by their usage in text (window size is five), and learns a vector for each concept such that concepts with similar co-co-occurrence patterns have similar embeddings. The result is a space in which concepts with *similar* embeddings have similar meaning and concepts with *dissimilar* embeddings have different meanings.

We learn embeddings of 100 dimensions, but the metric is robust to 100, 200, or 300 dimensions as well as to stochasticity. We capture the dominant meaning of a concept globally over time. (Although concepts may evolve over time, we use the globally dominant meaning of the concept because we also model uptakes of links globally, and modeling concept embeddings over time is computationally intensive; and suffers from data sparsity. Sensitivity analyses for one year (2000) provided very high correlations ($r = .931$) between global and time-dependent distal novelty scores.)

Having learned concept embeddings, we calculate how distant newly linked concept's embeddings are to one another using cosine distance (*35*) (see Table S4). We then average those scores for all novel links introduced in each thesis (*distal novelty*, Mean = .426; Median = .419; SD = .118). We validate these automatic measure of concept distance with expert human coders, finding moderate intercoder agreement between distal/proximal assignments to a random set of 100 links and three coders (average Cohen's Kappa = .46) and, together, coder assignments predict ~95% of the true distal links (i.e., distance score > .8). This validation further suggests that distal links are often between concepts from different fields or creative metaphors, and only a fraction of links between distal concepts are hard to interpret substantively (15-20%).

*Outcome variable–Careers*

To measure innovation reception, we study how innovations relate to two science career outcomes. The first is a conservative proxy of whether graduate students become research faculty after their graduation (*research faculty*, mean = .066). This outcome is measured as graduating PhDs who go on to become a *primary advisor* of other PhD students in the dissertation corpus. Ultimately, this captures who transitions from student to mentor at a PhD-granting US university, and who was able to secure a faculty job with a lineage of students. For those that graduated up until 2010 (i.e., the last graduating cohort we follow), we do consider whether they transitioned to faculty between 2010 and 2015. The second outcome is a more liberal proxy of career success that reflects whether graduating PhDs continue their career in research or not. To capture this, we match students to article authors in Web of Science (see Supplementary Text). The WoS database consists of ~38 million publication records and their associated meta-information from 1900 to 2017 (disambiguated authors, title, abstracts, etc.). The linkage across datasets allows us to follow students' ensuing careers and research output. Using the ProQuest-WoS link, we measure whether students publish academically at least once in the five years after obtaining their PhD *or* if they become research faculty, which we interpret as scholars who continue research endeavours (*continued research*: mean = 0.319). This metric captures a broader range of those who continue to pursue research: scholars who continue to pursue science at institutions that might not grant PhDs (e.g., liberal arts colleges, think tanks, industry jobs, etc.) or move internationally. Individuals from underrepresented groups might disproportionally move towards such institutions rather than US PhD-granting universities. Hence, examining both metrics indicates whether our results are robust to different academic strata.

*Main covariates*

The ProQuest dissertation data do not contain direct reports of student *gender* and *race* characteristics, but we identify the degree to which their name corresponds with the race or gender reported by persons with particular first (gender) and last (race) names. We compiled datasets from the US censuses (*38*) to predict race and from the US Social Security Administration (*39*) to predict gender. We matched these to data on N = 20,264 Private University scholars between 1993 and 2015. The Private University data contain race and gender information alongside scholar names, which allows us to train a threshold algorithm to estimate race and gender based on names. Using these thresholds, we classify advisees in the ProQuest dissertation data into one of three race categories and to assign a gender (*40*). The race categories are *White*, *Asian,* and *Underrepresented* *minorities*. Underrepresented minorities combines Hispanics, African Americans, Native Americans, and any racial categories not captured by the first three (see Supplementary Text). To further improve recall on genders and races, we focus on uncategorized genders and races and label them based on additional methods for gender (see *41-43*) and race (with full names: *44, 45*), thus combining the strength of several methods to help increase coverage and precision for gender and race labels.

We then measure the fraction of students in a discipline-year carrying the same gender or race – e.g., the percentage of women in Education in 1987 when a student is a woman, the percentage of underrepresented minority scholars when a student is an underrepresented minority, and so forth (*% Same-gender*, Mean = .576; SD = .180; *% Same-race*, Mean = .625; SD = .258). We also measure whether a student is part of an underrepresented gender or race in an academic discipline – i.e., whether a student is member of a group smaller than the largest group in a discipline-year (*Gender minority* mean = .336; *Racial minority* mean = .246, see Figure S4).

To model novelty, impactful novelty, and distal novelty, we use the percentage of same-gender/race and whether scholars are white/non-white to find to what extent innovation relates to different notions of group representation in science. We then model careers through minority status in disciplines (results are similar for binary gender/race indicators).

Note that the results here do not take into account cases of gender and race that were not classified according to these methods, although the gender and race distinctions such as shown in Figure 2C and F do not qualitatively change if we do include "unknown" genders/races in the analyses. Our main substantive conclusions/inferences are robust if we only consider those students whose names overwhelmingly occur within one rather than multiple races. Additionally, finer-grained notions of race or even degrees of identity association with gender or race may be desirable as an indicator. However, underrepresented races appear often in small proportions, which provide little statistical power in spite of likely sharing a common pattern of associations. As such, we render them into coarser indicators of "underrepresented racial minority." We recognize that in reality, individuals and names have varying degrees of gender and racial associations; as such our named-based metric is a simplified signal of gender and racial identity that may better capture how an individual is perceived by others, and can be only a coarse proxy for authors' self-identification with certain genders or races.

*Confounding Factors*

When dissertation metadata did not include a department, we identified academic discipline for theses filed with ProQuest through a Random Forest Classifier (RFC) based on a list of features from the dissertation with 96% precision ($N_{DISCIPLINE}$ = 84; see Supplementary Text). Dissertations that are filed to ProQuest contain meta-information about the *institution* where the doctorate was awarded. We classify the student into the first institution that students filed to ProQuest ($N_{UNIVERSITY}$ = 215). We infer the *graduation year* in which students obtain their doctorate as the year in which the dissertation was filed to ProQuest (Range = 1977-2015).

*Analytical strategy*

We model each of our dependent variables tailored to their statistical distributions. Scientific novelty (# new links) and impactful novelty (uptake per link), are right-skewed counts of events or rates. For these outcomes, we employ negative binomial regression analyses, where the over-dispersion in the outcomes is modeled as a linear combination of the covariates (*46*). Distal novelty is relatively normally distributed and we model it through linear regression. Academic careers as becoming research faculty (yes/no) and sustaining a research career (yes/no) are both binary outcomes, so we use logistic regression analyses for these (see Supplementary Text). The whiskers and shaded lines in Figures 2-4 represent upper and lower bounds of 95% confidence intervals, the *p*-values we report here are two-sided tests. Figures 2A, B, D, E, 3D, and 4 all represent average marginal effects considering all other values of the other independent variables. Figures 2C and F report the incidence rate differences between groups from the negative binomials regressions.

Apart from the main covariates, we include three sets of fixed effects in our models so as to better isolate our main predictors from confounding factors. We keep institution, academic discipline, and graduation year constant throughout. These fixed effects account for university differences in prestige and the resources they make available to students (*33*), the differences across academic fields and disciplinary cultures (*34*), and for "older" scholars that have had more time to make career transitions or to get recognized.

We weigh the data by the total number of doctorates awarded by an institution in a given year (see Supplementary Text) to account for possible selectivity between universities in years when filing their doctorates' theses in the ProQuest database and to render our results generalizable to the US scholarly population. These survey weights are based on the relative number of PhD recipients in the ProQuest data vis-à-vis the US PhD population per year for each university.

Finally, novelty (# new links) is modeled for students with non-missing values on all features (N = 808,375), impactful novelty (uptake per new link) is modeled for those with nonzero novelty *and* nonzero uptake given its best fit with the negative binomial model (N = 345,257), distal novelty is modeled for the students with nonzero novelty (N = 630,971). Careers are modeled for those where there are no constant successes/failures within the fixed effects, and for those who introduce at least one link (N = 805,236) or whose novelty is nonzero for impactful novelty (N = 628,738).

**Acknowledgements**

We acknowledge Stanford University and the Stanford Research Computing Center for providing computational resources and support that contributed to these research results. This paper benefited from discussions with Lanu Kim, Raphael Heiberger, Kyle Mahowald, Mathias W. Nielsen, Anthony Lising Antonio, James Zou, and Londa Schiebinger.

**Funding**

This paper was supported by two National Science Foundation grants [NSF #1633036 and NSF #1827477] and by a grant from the Dutch Organization for Scientific Research [NWO #019.181SG.005].

**Author contributions**

B.Ho., V.K., and D.M. conceptualized and designed the study. B.Ho., V.K. S.M.N.G., B.He., D.J., and D.M. analyzed the data. B.Ho. and D.M. did the writing.

**Competing interests**

The authors declare no competing interests.

**Data and materials availability**

The data used in this study were obtained according to protocol 12996, approved by Stanford University. We acquired written permission from ProQuest to scrape and analyze their dissertation data for scientific purposes. The full dissertation corpus can be requested via ProQuest (*20*) and the Web of Science can be requested via Clarivate Analytics (*36*). Code to replicate our key metrics is found on GitHub (https://github.com/bhofstra/diversity_innovation_paradox). Top terms from the K = 500 structural topic model that equally balances frequency and exclusivity is also found there.

# Supplementary Information for

## The Diversity-Innovation Paradox in Science

Bas Hofstra[1*], Vivek V. Kulkarni[2], Sebastian Munoz-Najar Galvez[1], Bryan He[2], Dan Jurafsky[2], & Daniel A. McFarland[1*]

[1] *Stanford University, Graduate School of Education,*
*520 Galvez Mall, Stanford, CA 94305, USA*
[2] *Stanford University, Department of Computer Science,*
*353 Serra Mall, Stanford, CA 94305 USA*

* To whom correspondence should be addressed: bhofstra@stanford.edu or dmcfarla@stanford.edu

**This PDF file includes:**

    **Supplementary Text**
        *Diversity and Innovation*
        *Measuring Innovation Through Citations, Keywords, and Text*
        *Structural Topic Models for Concept Extraction*
        *The PMI Score to Identify Meaningful Links*
        *Student Gender and Race*
        *Academic Discipline*
        *Population Coverage and Data Weights*
        *Inferential Models*
        *Linking ProQuest to Web of Science*

    **Figures S1 to S4**

    **Tables S1 to S4**

    **SI References**

**Supplementary Text**

*Diversity and Innovation*

Historically underrepresented groups come from distinct walkways, and have different experiences and perspectives than majority group members in science. Given this "outsider" vantage, underrepresented groups may perceive things differently from the majority group members, drawing relations between ideas and concepts that may have been missed or ignored. As such, they may be more likely to create novel connections between ideas in comparison with individuals who share the background and experiences of the traditional group already in place. Therefore, the inclusion of underrepresented groups in science may increase the variety of perspectives brought to bear on scientific research. Such intuitions align with recent work (*1-4*). For instance, Page (*1*) (see also Bell et al. (*2*)) reviews a large and growing body of evidence revealing how greater gender and racial diversity on teams creates a more heterogeneous pool of thinkers, and that in the long run (after some conflict), these groups are more innovative and outperform more homogenous groups. In the case of underrepresented groups – such as women and minorities – they may bring a perspective and set of concerns that are missed by the majority groups present in science (who were the "default" from the outset), and therefore, they introduce heterogeneity to the collective thought process. This heterogeneity generates what he calls "diversity bonuses" to improved problem solving, increased innovation, and more accurate predictions.

*Measuring Innovation Through Citations, Keywords, and Text*

Prior researchers have studied citations or keywords to understand scientific innovation. For instance, some prior work (*5*) regarded novel recombination of bibliographic sources to be a sign of innovation. Here, we extend prior work by using recombinations of concepts used in scientific text, thus likely maintaining references to the explicit meaning of said concept combinations (*6*).

Keywords are an alternative to citations. They constitute "plausible building blocks of content" (*7*), and get at taxonomic aspects of scientific knowledge. Prior work used keywords to identify where innovation arose from subfield integration (*8*). An issue with keywords, however, is that it is difficult to ascertain whether they classify the general topic of a paper or refer to specific contents and innovations contributed by it. Researchers, and often editorial teams, assign keywords to optimize indexing and retrieval (*9*). The use of keywords then begs the question of whether they locate innovation in a research article or in its classification.

As an alternative to keywords, prior work used chemical entities from annotated MEDLINE abstracts as their units for innovation (*7*). By extracting chemical entities from abstracts, this work overcome potential confounding with classification dynamics. Yet, the study of chemical entities is highly specific to one field: chemistry. As such, scholars acknowledge, "new methods should be developed for mining building blocks with finer granularity" (*7*: 901). Our analysis of novel recombinations of concepts in documents overcomes the issues of citations and keywords and thus elaborate and extend the research program on innovation.

There are at least two more advantages to measuring innovation and impact with the language of PhD recipients in dissertations vis-à-vis citation records of scholars in journals. First, language metrics are relatively unaffected by academic search engines, journal guidelines, or differences in indexing across corpora, or by the variety of reasons as to why scholars cite others' work (*6, 10*). As such, we detect signals of innovation that may otherwise be hard to trace and which are insensitive to potential biases resulting from corpora that unjustifiably exclude citations in other academic fields. Second, our corpus captures a near-population of scholars' early texts and does not discriminate by prioritizing some academic fields at the expense of others. As such, the language and innovations of slower, book-oriented science (e.g., History), medium-paced, publication-oriented science (e.g., Sociology), or faster, proceedings-oriented science (e.g., Computer Science) are all represented and measured in our corpus.

Finally, a potential drawback of our universe (ProQuest dissertations) is that the link introductions we identify in ProQuest dissertations might have arrived earlier in other corpora (e.g., peer-reviewed journals, or even fiction). However, it provides (at the very least) unique insight into which dissertations are novel compared to others dissertations and, thus, which students are competitive vis-à-vis others with their earliest innovative sparks in the knowledge they produce.

*Structural Topic Models for Concept Extraction*

*Structural Topic Models*

To identify scientific novelty in concept use, we first fit Structural Topic Models (STMs) (*11*) where we model the prevalence of topics in dissertation abstracts (~1.2 million) as a linear function of the year in which scholars obtained their doctorate. Structural topic modelling is an unsupervised learning technique that represents texts within a corpus as a mixture of latent thematic dimensions without *a priori* knowledge of what these dimensions might be. STMs rely on co-occurring words within documents. In an iterative process, this kind of model draws samples from a corpus to derive a series of topics – i.e., weighted sets of co-occurring words in a text. The outcome of this process is twofold: (a) the model arrives at the set of topics best suited to explain the thematic dimensions of a corpus of texts; and (b) the model produces an optimal representation of every document as a mixture of topics.

More formally in a topic model, a given topic $k$ is associated with a probability mass function $\beta_k$ over a given vocabulary $V$ (*12*). In every document $d$, the model draws a topic for each word position $n$ from a multinomial based on a global prior distribution over topics, $\theta_d$. Then, the model draws the observed word $w$ for position $n$ from a multinomial based on $\beta_k$. The distribution $\beta_k$ associated with topic $k$ controls the probability of drawing the $v$-th word in the vocabulary for topic $k$ (*12*). The model learns the $\theta$ and $\beta$ distributions via variational expectation-maximization (see *13, 14*). For the purpose of this study, we can think of the distributions $\beta_k$ as distinguishing important words in the vocabulary with respect to topic $k$. STMs are a particular kind of topic model that allows us to include additional information into the model (*12*); namely, we estimate $\theta$ as a function of the year of publication of the dissertations. Specifically, we allow topics to be more or less prevalent over time. We do this by modelling the prevalence of topics in dissertation abstracts as a linear function of the year in which scholars obtained their doctorate. We found that including publication metadata in $\theta$ had little impact on the $\beta$ distributions we use to extract documents. Trying to model $\beta$ distributions directly as a function of year of publication was computationally intractable. Further, we discovered that STMs and simpler LDA topic models produced very similar $\beta$ distributions. We opted to keep the STMs as they introduce more information without a loss in the quality of the topics or their interpretability.

In our corpus, topics refer to areas of scientific research and discourse. We extract terms that STMs identify as the most distinct and heavily used *within each research area.* We contend that scientific innovation involves novel combinations of such terms. The affordance of STMs in comparison to simpler concept extraction strategies –i.e., choose the top n TF-IDF weighted terms– is that it allows us to extract terms that play a significant role in an underlying thematic structure.

We mention "best-suited" topics and "optimal" document representations because STMs, like other mixture models of its kind, allow for the validation of different numbers of possible latent dimensions or themes. Here, we fit STMs within a range of a set number of topics [K = 50-1000], with incremental steps of 50 (and steps of 100 when K > 600 to save computing time). Internal and external validation indices show that the optimum of the number of topics is at approximately K = 400-600 topics. In the main text of the study, we have presented results for K = 500. This means that we used the weights on the vocabulary from an STM with 500 topics to extract the concepts that best describe the latent dimension in the corpus. Namely, the extracted concepts belonging to the highest FREX-score terms of each topic (detailed below). However, our results remain robust under alternative specifications for concept extraction (leaning towards either frequency, exclusivity, or balancing both equally) and for a range of K (for 400, 500, and 600) (see Table S1). Next, we first detail how we preprocess the data and arrive at K = [400-600] based on several fit metrics, and then outline the concept extraction using FREX.

*Preprocessing Texts and Fitting STMs*

We preprocess the data by the following steps. We remove stand-alone numbers, punctuation, English stop words, and special characters from the text. However, we keep numbers belonging to terms such as molecules (e.g., $H_2S$), which might refer to substantive concepts. We then stem the words using the Snowball algorithm and remove those tokens that only appear once across all documents. We extract n-grams for sequences of words that occur more frequently than by chance using El-Kishky et al.'s method (*15*). We then fit STMs at K [K = 50-1000] in incremental steps of 50 (and steps of 100 when K > 600 to reduce computing time) by training each for 20 epochs.

*Internal validation*

We then *internally* validate the models to find out what number of topics retrieves the most-discriminant latent thematic dimensions; which is equivalent to finding the dimensionality reduction solution that retains the most information about the corpus. To do so, we consider both the *coherence* and *exclusivity* (*11, 16*) of the topics produced by models at different values of K.

The coherence of a topic assesses its internal consistency. Semantic coherence is obtained by calculating the frequency with which high-probability words within a given topic co-occur in documents. The most-probable words in a highly-coherent topic tend to appear together in documents. Conversely, a low-coherence topic comprises high-probability words that appear in isolation from each other. It would be difficult to argue that a low-coherence topic is of much use in representing documents, since it can appear in multiple documents with very different terms.

Assessing topics solely on their semantic coherence is not enough, since this measurement can be trivially maximized by reducing the number of topics. For instance, if we had a single topic, high-probability terms would co-occur by construction. Similarly, a topic that comprises very common words of a topic (e.g., data, study, etc.) will appear to be very coherent since these terms co-occur in most documents by convention. Therefore, as a complement to semantic coherence, we want our model to produce topics that have very distinct high-probability terms; that is to say, we want topics with high exclusivity. Exclusivity measures the extent to which words within a topic are distinct from the words in other topics. There is a trade-off between a topic's exclusivity and semantic coherence – i.e., overall high-probability words tend to drive very coherent topics, since they are likely to co-occur; but these words also tend to co-occur with the terms from many topics, and so they drive low exclusivity topics. Given this trade-off, we explore the solution space along values of K looking for the model where both exclusivity and coherence plateau and do not improve nor decrease with a lower or higher number of topics, thus providing us with a potential limit for K. Figures S1-A and S1-B shows that this limit is likely to be in the range of K = 400-600.

*External validation*

In addition to internal validation, we also employ external validation. To this end, we compare the distance between documents based on an STM with a given K with the document distances based on author-provided keywords and fields. We use the academic fields and keywords that students file with their dissertations. We draw a random sample of 1000 documents that remains constant across values of K, and compute the cosine similarity between document pairs in this sample based on the documents' topic mixtures. In so doing, we leverage that all document pairs are comparable in vector θ, which represents any given document as a probability distribution over all topics. We then consider any given document pair to be related if their cosine similarity is greater than the median similarity in the sample. For the field and keyword relations between documents, we consider whether bigrams (fields + keywords) occurring *within* a document co-occur *between* two documents; when this is the case, we render these documents related.

We represent the relations described above as two document-to-document networks, one STM-based and one bigram-based network, and study their overlap. We are interested in four kinds of comparisons at the level of document dyads, which we can picture as a two-by-two matrix where the rows indicate if a document dyad appears in the STM-based network (Yes/No) and the columns indicate if the dyad appears in the bigram-based network (Yes/No). Given the comparisons of interest, we compute the Matthew correlation coefficient, which measures the overlap at the dyad level between the STM and bigram networks. An advantage of the Matthew correlation metric is that it accounts for overlap on true negatives (i.e., when a document dyad does not appear in either the STM or the bigram network). The Matthew correlation coefficient is defined as follows: *Matthew correlation* $= \text{TP} \times \text{TN} - \text{FP} \times \text{FN} / \sqrt{(\text{TP}+\text{FP})(\text{TP}+\text{FN})(\text{TN}+\text{FP})(\text{TN}+\text{FN})}$, where T and F define true and false, and P and N define positives and negatives. Figure S1-C depicts the result of the correlations between keyword and STM relations. We find that the curve follows a similar trend compared to the internal validity metrics. There is a decrease as K moves beyond 500, providing some external validation with user-labeled information that the number of topics seems to optimize around K = 500.

*Consistency*

Additionally, we study the consistency of topic assignments across the range of K [50-1000] – i.e., whether the topics retrieved at one value of K are informative of the topics obtained at another value. To this end, we first classify all documents by their highest-proportion topic at each value of K. This step results in a set of classification schemes, one scheme for each model with a different value of K. We then compare the classification schemes of consecutive models (i.e. the document classification under K = 50 compared to the classification under K = 100) using the Fowlkes-Mallows index (FM). The Fowlkes-Mallows (FM) index measures overlap between two distinct clusterings of the same data set. FM is part of a family of indices for external clustering validation, such as the Jaccard coefficient and the Rand statistic, that use the agreements and/or disagreements of the pairs of data objects in different partitions (*17*). This measure is the geometric mean between precision and recall and is bounded between 0 and 1; higher values indicate greater agreement between two partitions – i.e., implying higher similarity in how the two partitions are clustered. There are external clustering validation metrics for multi-labeled corpora, like the document-topic matrices produced by STMs (*18*), but we use the FM index for simplicity, as we want to describe the alignment of multiple STMs at different values of K. It shows us at which K the overlap plateaus to pinpoint our number of topics.

In Figure 4-D we describe the *rate* at which the overlap between classification schemes vary when comparing each model with K topics to the immediate prior model with smaller K. We see relatively high values of consistency with a gradually growing curve, which suggests that classification schemes are more similar at the higher end of values of K. The range of K suggested by FM is in line with the previous measures: we see a steady rise and somewhere between K = 400 and K = 600 it stabilizes and only a gradually improvement afterwards. Raw FM scores suggest that more than two-thirds of document-to-topic assignments are stable from K = 400.

*The "Right" K*

Finally, we emphasize that we do not use the "right" K, as that would imply that we are perfectly aware of the topic (and, hence, scientific) universe. We use K = 500 in the main text as the metrics all seem to plateau around that value. However, if we choose K = 400 or K = 600 and measure concept/link introduction and uptake in a similar way (using low, medium, and high FREX-weight), our results do not qualitatively change. The "right" K – if one is to interpret that as the set number of scientific topics at the specialization within disciplines level – likely is somewhere between 400-600. A benefit of our approach, and what our associated results show, is that the key results stay mostly qualitatively similar whichever K (400, 500, or 600) we choose.

*Concept Extraction With FREX*

Using the STM output, we then obtain the most-frequent and most-exclusive terms within a given topic. The most-frequent terms reflect general language present in many of the topics (e.g., "data," "analyze," "study," etc.), whereas the most-exclusive terms may be too idiosyncratic to be informative in and of themselves (e.g., "eucritta melanolimnete," "periplanone b," etc.). Concepts that are both common and distinctive balance generality and exclusivity. To get at this, we extract concepts on the basis of their FREX score (*19*), which compounds the weighted frequency and exclusivity of a term in a topic. Here, we explore three weighting schemes: equally balancing frequency and exclusivity (50/50), attaching more weight to frequency and less to exclusivity (75/25), and attaching more weight to exclusivity and less to frequency (25/75). We then extract the top-500 FREX-words per topic – K = [400-600] with incremental steps of 100 – and measure our innovation variables for all three K's and three FREX weighting schemes (i.e., nine scenarios in total). The more-frequent semantic space defines the more-standard scientific vocabulary, and the more-exclusive semantic space is more idiosyncratic indicative of non-standard concept usage. Sensitivity analyses provide robust results across the scenarios for novelty, impactful novelty, and recognition (see Table S1). For the results depicted in the main text, we report the scenario where frequency and exclusivity are equally balanced at K = 500.

*On Analyzing Abstracts Versus Full Texts*

We analyze dissertation abstracts based on the conjecture that abstracts are a good approximation of the knowledge and concepts that populate full texts. Prior work consistently shows that this conjecture is a reasonable one, as abstracts provide a clean, uncluttered synthesis of the full text. Prior work suggests that the goal of abstracts is to summarize and emphasize a paper's key contributions (*20*). Empirical work observes that abstracts provide sufficient syntheses of concepts, tables, graphs, and topics in papers (*21-24*). Pragmatic arguments in favor of using abstracts is that the use of full text is highly restricted by its general inaccessibility, biased sampling, poor scalability, and high demand on computational resources for large corpora. In contrast, abstracts are easier to obtain and typically demand far fewer computational resources. Additionally, with the use of full text come some theoretical difficulties. For instance, if we study concept co-occurrence in full text, at what distance do concepts need to co-occur in order to render the co-occurrence substantively meaningful? In the same text, section, paragraph, or sentence? Co-occurrences in abstracts are far more likely to be substantively meaningful as abstracts only cover ~10 sentences. Finally, our main results would only qualitatively change if numerical minorities write abstracts that are inherently different compared to those written by majorities. Given the general goal of abstracts – i.e., summarizing main contributions and findings (*20*) – we assume that the retention of innovations in abstracts versus full text is not higher (or lower) for numerical minorities vis-à-vis majorities.

### *The PMI Score to Identify Meaningful Links*

The significance score (*25*) for links is defined as follows, given a concept link $L = (a, b)$ we compute such a score as:

$$PMI(L) = log10\left(\frac{Pr(a,b)}{Pr(a) \times Pr(b)}\right), \qquad (1)$$

where Pr(*a, b*) is the likelihood that concept link a–b occurs, Pr(*a*) is the likelihood of concept a and Pr(*b*) is the likelihood of concept *b*. "Good" links will then result in a high PMI scores – significantly more likely to occur than chance. We then filter for spurious recombinations using a rank-based cutoff based on the *PMI* score. To ensure sufficient power for computing *PMI*, we only consider those links where individual terms occur in at least 10 theses. We consider the top 10 million links, so as to have ample opportunity to introduce "novelty" while simultaneously removing obviously meaningless links. This is achieved by setting a cut-off on the *PMI* score for link introductions.

*Student Gender and Race*

The ProQuest dissertation corpus (*26*) does not contain records of gender and race of students that filed their theses. Therefore, we predict the race and gender of students based on their first (gender) and last (race) names (*27*). For race, we compiled US Census data of 2000 and 2010 (*28*). These censuses show relative frequencies of racial backgrounds of last names that occur more than 100 times (N = 167,409 distinct last names that cover > 95% of the US population). For instance, it shows the fraction of individuals who carry the last name "Jones" whom are white. The correlation in racial background percentages of overlapping names (N = 146,516) in both censuses is .99. For gender, we compiled data of the US Social Security Administration (*29*). This corpus shows the fraction of girls and boys among the top 1000 first names from people born from 1900 to 2016 (N = 96,122 distinct first names that occur at least five times) – e.g., the fraction of girls named "Jane."

We matched distinct last names of the censuses to the last names (up to the first space or hyphen) in data from Private University where we are aware of self-reported race ($N_{total}$ = 24,150; $N_{match}$ = 20,264 [83.9%]). We matched all distinct first names of the social security data to the first names in the Private University data where we are aware of self-reported gender ($N_{total}$ = 35,469; $N_{match}$ = 31,026 [87.5%]).

An algorithm automatically traced which thresholds of the fraction of the last- and first-name carriers' race and gender yield the highest possible correlations between real and assigned gender or race. It did so by correlating self-reported gender and race with all permutations of the thresholds in steps of 1 percent. A threshold where at least 71.45% of the first-name carriers are female to assign students to a female gender provided the highest correlation between self-reported and assigned gender ($r = .91$). In order to identify the gender signal of those names that are not classified according to this prior classification scheme, we employ the Genderize.io classification scheme (e.g., see *30-32*) (agreement between our and the Genderize.io classification is > 95%). We arrive at ~8.5% of cases with unknown genders.

Additionally, the highest correlations for race were .83 (white, 12,929 of 13,197 identified correctly [97.2%]), .93 (Asian, 5,079 of 5,436 identified correctly [93.4%]), .73 (Hispanic, 698 of 992 identified correctly [70.4%]), and .25 (African and Native American, 63 of 639 identified correctly [9.9%]). Using these thresholds, we classify students into a racial background and gender. If students are classified into multiple races given our thresholds, we use a decision rule; (1) when a student was classified into the African and Native American or any other category, we classify them as African and Native American; (2) when a student was classified into the "Hispanic" and "white" or "Asian" category, we classify the student as "Hispanic"; (3) when students were classified into the "Asian" and "white" category, we classified the student as "Asian" category. Finally, if the thresholds did not classify a student into a category, we used a majority rule to categorize the student into a race. For instance, when "Yao" does not meet a threshold while most individuals named "Yao" are in fact Asian we classify these as "Asian."

The fraction of correctly identified in the "African and Native American" category is low. We found that these students are predominantly labelled under "white" (528 white out of 639 Other Race). We therefore incorporate a second method that utilizes the sequence of characters for classification of race using names (*33*). Specifically, we utilize their method using full names in the Florida voting registration data (*34*). The precision of this method is especially high for Hispanic and African American names – .83 and .74, respectively – so patching our classifications with theirs combines the strength of our "white" and "Asian" and their "Hispanic" and "African American" classifications. If the probability of a certain name being Hispanic or non-Hispanic Black is higher than .6 using their method, we label those cases as such. We ascertain that those cases are highly likely to belong to those categories. Additionally, if our classification yielded an "unknown" case, if the probability that a name is Hispanic or non-Hispanic Black is higher than .3, *and* that probability is twice as high as the probability that a name is "White," we label those cases as Hispanic or African American. This filters the classification for very low probabilities, while simultaneously being confidant that those names significantly differ and do not have a clear signal for whites. Finally, if cases are still unknown, and the probability of a name being Asian or White is higher than .5, we label those cases as such. Our number of cases with unknown race is ~10.8%. Nonetheless, to ensure that remaining errors in our classification of race by name do not affect our results, we run a sub-analysis using only the highly certain cases for inferring race. The results using this smaller but higher-precision dataset are qualitatively similar to the ones presented in the paper.

*Academic Discipline*

Some theses do not identify the department from which they got their degree. To infer this, we first extracted theses *with* department degrees in ProQuest dissertations. Each department was then semi-manually canonicalized to a National Research Council (NRC) department. Given that there are many spelling mistakes, a fuzzy string matching was used to match the ProQuest department with the actual listed NRC departments based on a 90% string similarity (a manual analysis showed 100% accuracy). For the frequent department names that matched around and 70-89% to an NRC department, each canonicalization from ProQuest to NRC were manually verified (while rejecting those that were invalid). All dissertations whose department name could not be mapped to an NRC department had their department inferred as if it had not been listed. We used the successfully matched dissertations with an NRC department (N = 178,511) as a ground truth. Next, we trained a Random Forest Classifier (RFC) based on a list of features from the dissertation; binary features for whether the dissertation was listed with an NRC subject category, binary features for whether the dissertation was listed with ProQuest subject category, all keywords used for the dissertation, the topic distribution of the dissertation abstract using a 100-topic Latent Dirichlet Allocation model, the average Word2Vec word vector for each of the (1) keywords, (2) ProQuest fields, (3) NRC fields, and (4) title, and the degree-granting university. The RFC infers department degree with 96% precision ($N_{DISCIPLINE} = 84$).

*Population Coverage and Data Weights*

During the study period (1977-2015) approximately 1.2 million doctorates were awarded in total. This suggests that the ProQuest data cover approximately 86% of the total number of US doctorates over three decades. If we plot the ProQuest database and the population of awarded doctorates in the US over time, the trends are highly similar. In our inferential analyses, we weigh the data from 1982 to 2010 by the total number of doctorates awarded by an institution in a given year to account for possible selectivity between universities in years in filing their doctorates' theses in the ProQuest database. To do this we calculate for each distinct year-university combination (e.g., at Harvard University in 1987) the number of PhD recipients and divide this number by the total number of PhD recipients in the ProQuest data, 1982-2010. This yields the relative number of PhD recipients in the ProQuest data per year for each university. We repeat this calculation for the *total* PhD recipients according to the data from the National Science Foundation. We then divide the relative number of PhD recipients for the university-year combinations in the ProQuest data by the relative number of PhD recipients for the university-year combinations in the census to obtain our data weights. We use these weights as survey weights in our inferential analyses. We use Stata 13 for the inferential analyses in the paper and to compute average marginal effects shown in Figure 2-4.

*Inferential Models*

Analytically, our models take the following forms:

$$Pr(Y = y_j \mid \mu_j, \alpha) = \frac{\Gamma(y_j+\alpha^{-1})}{\Gamma(\alpha^{-1})\Gamma(y_j+1)} \left(\frac{1}{1+\alpha\mu_j}\right)^{\alpha^{-1}} \left(\frac{\alpha\mu_j}{1+\alpha\mu_j}\right)^{y_j}, \quad (2.1)$$

where

$$\mu_j = exp(\beta_0+\beta_1 X_j+\ldots+\beta_k X_j), \quad (2.2)$$

$$Y = \beta_0+\beta_1 X_j+\ldots+\beta_k X_j + \varepsilon, \quad (3)$$

$$Pr(Y \neq 0 \mid X_j) = \frac{exp(\beta_0+\beta_1 X_j+\ldots+\beta_k X_j)}{1+exp(\beta_0+\beta_1 X_j+\ldots+\beta_k X_j)}. \quad (4)$$

Equation (2) models the expected count/rate of link introductions (novelty) or uptake per new link (impactful novelty), equation (3) models the average distality of the introduced links by students, and equation (4) models career success as becoming a faculty researcher or sustaining a research career. All equations (2)-(4) are for individual student $j$. In these models, $\beta_0$ represent intercepts and $\beta_1 X_j+\ldots+\beta_k X_j$ represent our vector of covariates from the first to the $k^{th}$ variable that predicts the outcome $Y$. Variables included in this vector are our main predictors (e.g., indicators for gender and race representation) and the confounding factors (institution, discipline, and year).

Uptake per new link (impactful novelty) is a non-integer rate instead of an integer event count. An occasional method of modelling non-integers is to offset the negative binomial regression with logged independent variables. Here, we do so for the number of new links when we model uptake per new link so as to interpret coefficients of other independent variables as rate increases or decreases (*35, 36*). A (simplified) example is an expected count $\mu_x$, where $\mu_x$ is dependent on some covariate *X*, so that $log(\mu_x) = \beta_0 + \beta_1 X$. If $t_X$ would then indicate exposure (or offset), then $log(\mu_X / t_X) = \beta_0 + \beta_1 X$ models an expected rate (count divided by exposure) and this is analytically equal to $log(\mu_X) = \beta_0 + \beta_1 X + log(t_X)$. Hence, we include a logged offset variable $t_X$ in the form of logged number of new links. As such, we are able to model uptakes per new link as non-integer rates.

*Linking ProQuest to Web of Science*

We attempt to link each student in the ProQuest corpus to their corresponding identity in two sets of publication corpora from the Web of Science (WoS) obtained from Clarivate Analytics. The first set contains publications from 1900 to 2009 (~22 million) and the second set contains publications from 2009 to 2017 (~16 million). The matching process between ProQuest dissertations and both WoS corpora relies on substantial meta-data in each of the three data sources.

The pre-2009 WoS data does not contain canonical author identifiers with high precision so we use a disambiguated author cluster (*37*), which contains groups of publication records in WoS estimated to be authored by the same person with substantial certainty (83%). The post-2009 does contain disambiguated authors by Clarivate Analytics with substantial accuracy post-2009, but with poor accuracy pre-2009. In order to make optimal use of both disambiguated datasets, we needed to reconcile the pre-2009 clusters and the post-2009 clusters. Hence, the goal is to link these two author-disambiguated datasets so as to benefit from the high accuracy from both datasets across the whole time range and increase coverage throughout. We pinpointed which author-clusters in the pre-2009 set were which clusters in the post-2009 set. We generated a link between the pre-2009 and the post-2009 author clusters, indicating that both clusters are the same author, if any of the following conditions were met, in addition to sharing a full name: 1–75% of the pre-2009 cluster articles are a subset of the post-2009 cluster articles; 2–There is at least one matching email address between an old cluster and a new cluster. Once these rules were applied, we finished cluster linking by manually checking and verifying a random sample of entries, in addition to automated verification of linking rules being followed on a larger random sample. The method above is conservative in its creation of links as a result of the strictness requiring a 75% match in order to link. This approach prioritizes the reduction of mistakenly-linked clusters at the expense of undiscovered linkages. Precision of the line-up between the two sets is 97%, which we inferred from a set of online, self-labeled publications by scholars that Clarivate Analytics provided (ResearcherID).

In turn, matching between WoS (linked pre- and post-2009) and ProQuest dissertations follows a multi-step sieve process, where scholar matches are evaluated using multiple successive criteria starting with the highest-confidence first: (1) number of article co-authorships with a known advisor or advisee, (2) number of articles where the WoS author is at one of the same institutions from the ProQuest data (as an advisor or advisee), (3) number of article keywords matching those from their dissertation keywords, (4) minimum string similarity of the authors' names (as reported for each article) with the name in ProQuest, and (5) textual similarity of the articles' abstracts and titles with the dissertation abstract. For naming similarity, our method is robust to minor typographic errors in names (as ProQuest information is manually entered) and to recognized naming variants (e.g., Dave or David) and abbreviations in the first and middle names of the individuals. This entire matching process amounts to a maximum bipartite matching of the ProQuest and WoS authors, ensuring that one author from either side is never linked to more than one author on the other side.

As this matching process could potentially be noisy, we take additional steps to heuristically reduce the potential for mismatches. First, we restrict WoS matches to only those individuals whose publication history is similar to their graduation date; this restriction excludes matching those individuals whose nearest publication date is 15 years after or 10 years before graduation. Second, we avoid matching individuals where the bulk of their publication occurs before their graduation, except in the case where there is additional evidence to support the matching from co-authorship with their advisor. Third, we avoid matching individuals whose only evidence for being the same person is their name similarity and a textual similarity between their dissertation and the articles (e.g., no evidence of being at the same institution where they would have graduated or advised students).

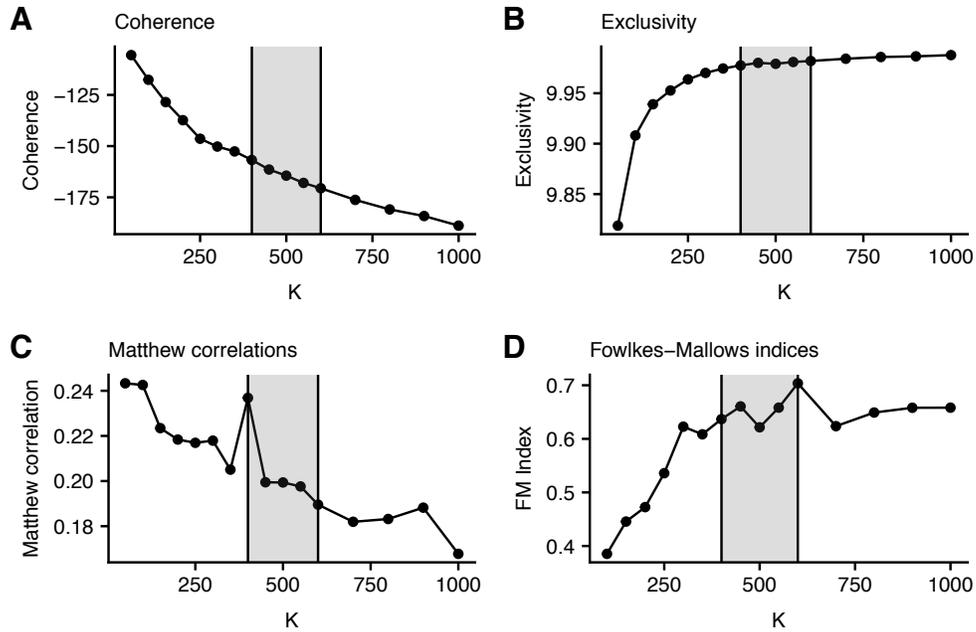

**Figure S1.** *Internal and external validity and coherence for structural topic models.*
*(A-D) We highlight the range of K we use (K = 400-600). (A-B) Values of coherence and exclusivity across a range of K. With a rising number of topics exclusivity increases but plateaus at approximately K = 400, while coherence decreases somewhat continuously, although less steep from K = 250. (C) Matthew correlations between external relations between documents and keywords and relations between documents derived from the topic models. The correlations spike at K = 400 and stabilize thereafter. Robustness analyses for K = 450 yield the same results as the analyses for K = 400. (D) Fowlkes-Mallows indices indicating overlap of topic-assignments for consecutive K's. The Fowlkes-Mallows correlation plateaus from approximately K = 300 and onwards, with a spike at about K = 600.*

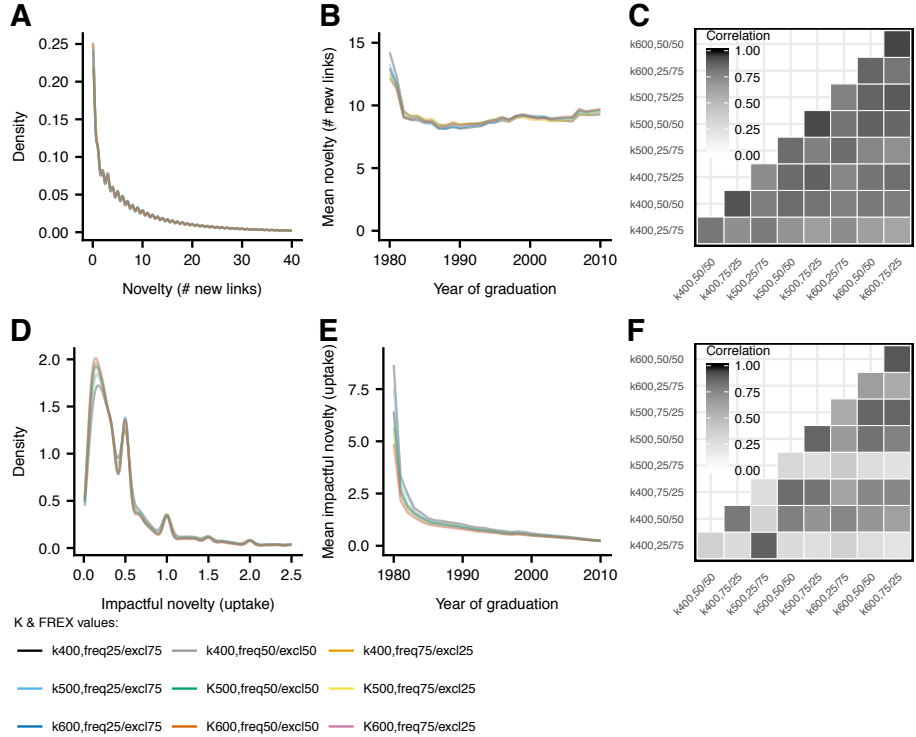

**Figure S2.** *Distribution of novelty and impactful novelty.*
*(A) Density distributions of novelty (# new links) for a different number of K and difference scenarios for FREX (low, medium, or high frequency). Despite absolute differences, the distributions are qualitatively similar. (B) Mean novelty over time. The figure suggests that the "stable" novelty starts at approximately 1982. The main paper analyzes the data from that point onward. (C) High correlations between the different novelty scenarios. (D) Density distributions for impactful novelty (uptake per new link), again suggesting similar distributions across the K and FREX scenarios. (E) Mean impactful novelty over time, suggesting that the "stable" impactful novelty starts at approximately 1982. (F) Relatively high correlations between the different scenarios for the measure of impactful novelty.*

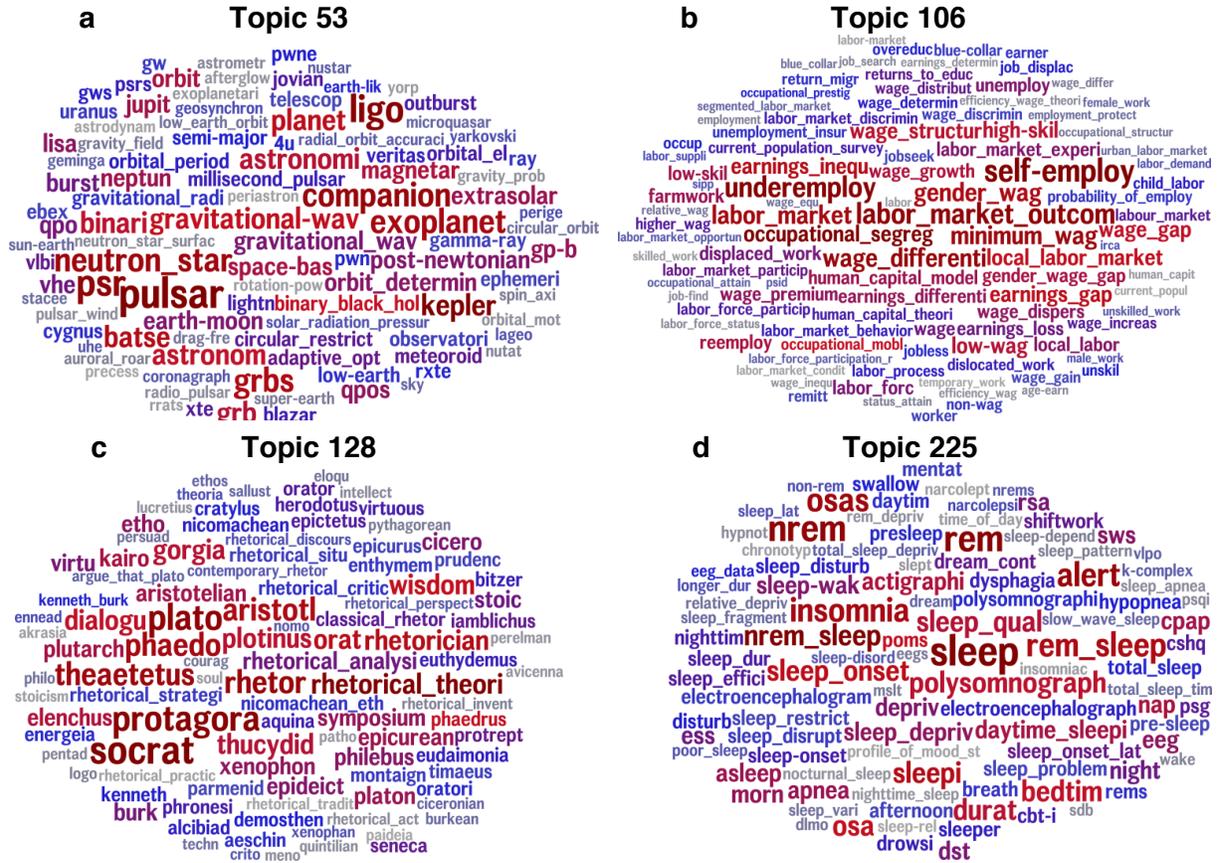

**Figure S3.** *Exemplary topics and their extracted concepts using FREX.*
*(A-D) Concepts from a small selection of extracted topics in our K = 500 topic model where we equally balance frequency and exclusivity. Topics are research areas and discourse themes characterized by set of co-used terms, some of which are more salient to the latent themes than others. (A) Students engaging in this topic are writing about astrophysics. (B) Students engaging in Topic 106 are writing about labor economics and income. (C) Students engaging in Topic 128 focus on rhetoric and classic Greek philosophers (with some exceptions). (D) Students engaging in Topic 225 are writing about sleep patterns and issues surrounding it. This is a small selection here, but the full set of topics is available online ([https://github.com/bhofstra/diversity_innovation_paradox](https://github.com/bhofstra/diversity_innovation_paradox)).*

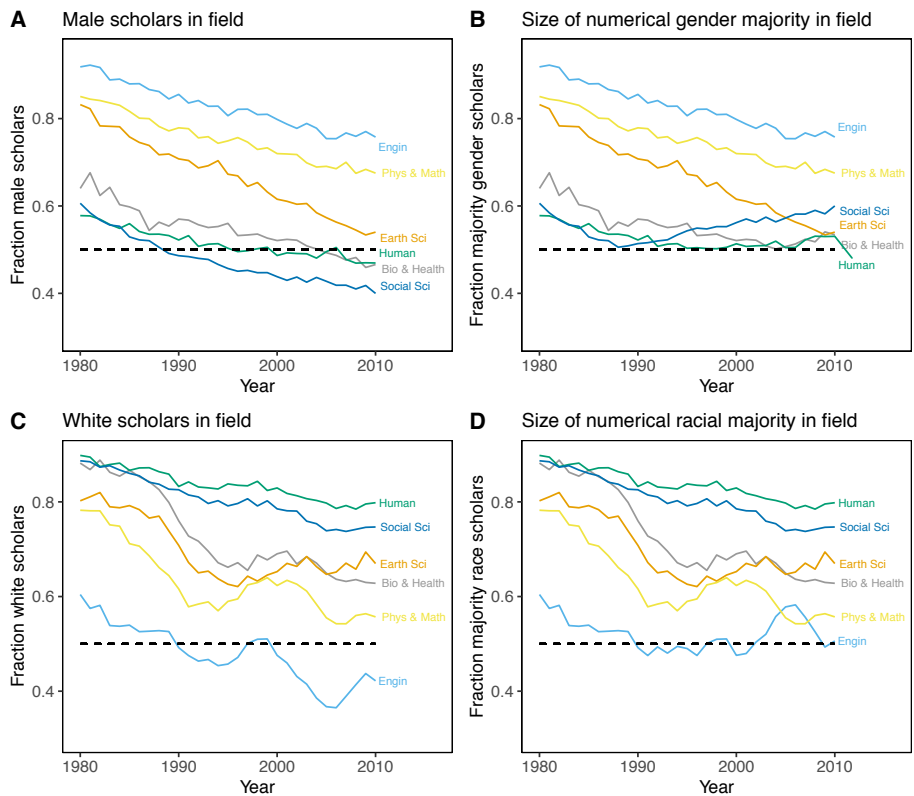

**Figure S4.** *Gender and racial representation of students in academic fields over time*
*(A-D) We aggregate the disciplines into broader academic fields for a depiction of minority statuses. Across all fields, women and non-white students are numerical minorities and keep that status very frequently. (A) Women become numerical majorities (and men minorities) when the fraction of male students drops below .5 (e.g., Social Sciences > 1990). (B) Depicting the size of the numerical gender majority. Women become majorities in case where the fraction of men drops below .5 in panel A, which happens only in few cases. (C-D) In very few cases do non-white students become numerical racial majorities (i.e., only if white < .5). However, becoming a numerical racial majority is not a given when white < .5, as there are more than two racial groups – i.e., whites (or another group) might still be majorities if the remaining fraction is split into several smaller nonwhite subgroups. Note that only in certain years do non-whites become a numerical majority in engineering.*

**Table S1.** *Descriptive statistics of the number of concepts in a dissertation abstract.*

|  | Mean | SD | Median | Minimum | Maximum |
|---|---|---|---|---|---|
| Overall | 56.50 | 19.44 | 57 | 0 | 356 |
| By race |  |  |  |  |  |
|   URM | 57.14 | 19.34 | 57 | 0 | 356 |
|   Asian | 56.69 | 19.12 | 57 | 0 | 219 |
|   White | 56.27 | 19.56 | 56 | 0 | 331 |
| By gender |  |  |  |  |  |
|   Women | 57.22 | 19.39 | 57 | 0 | 265 |
|   Men | 56.02 | 19.53 | 56 | 0 | 356 |
| By field |  |  |  |  |  |
|   Biology & Health | 60.02 | 16.77 | 60 | 0 | 248 |
|   Earth Sciences | 57.63 | 17.96 | 59 | 0 | 183 |
|   Engineering | 54.48 | 18.86 | 54 | 0 | 356 |
|   Humanities | 66.93 | 22.86 | 69 | 0 | 331 |
|   Physical Sciences | 50.16 | 19.38 | 51 | 0 | 219 |
|   Social and Behavioral Sciences | 54.43 | 18.03 | 54 | 0 | 265 |

**Table S2.** *Sensitivity analyses across K and FREX scenarios mostly show a similar pattern of results.*
*We find a qualitatively similar pattern of results across our K and FREX scenarios and this shows that most of our main results are insensitive to the way we extract concepts – i.e., weighing more to frequency or exclusivity – despite that the quantitative correlations might vary across scenarios. "Yes" in the table below indicates a statistically significant effect (i.e., one-sided p-value < .05) Note that we especially find some variable results in the lower K (400) or very high exclusivity scenarios. This likely results from our conservative filter to detect spurious links that we describe in the Materials and Methods. We present the "middle" scenario as the main one in the paper (K500, freq50/excl50).*

| | K400, freq25/excl75 | K400, freq50/excl50 | K400, freq75/excl25 | K500, freq25/excl75 | K500, freq50/excl50 | K500, freq75/excl25 | K600, freq25/excl75 | K600, freq50/excl50 | K600, freq75/excl25 |
|---|---|---|---|---|---|---|---|---|---|
| **Novelty (# new links)** | | | | | | | | | |
| % Same-gender ↓ # new links | Yes | Yes | Yes | Yes | Yes | Yes | Yes | Yes | Yes |
| % Same-race ↓ # new links | No | Yes | Yes | Yes | Yes | Yes | Yes | Yes | Yes |
| Women ↑ # new links | Yes | Yes | Yes | Yes | Yes | Yes | Yes | Yes | Yes |
| Non-white ↑ # new links | Yes | Yes | Yes | Yes | Yes | Yes | Yes | Yes | Yes |
| **Impactful novelty (uptake per new link)** | | | | | | | | | |
| % Same-gender ↑ uptake per new link | No | No | Yes | No | Yes | Yes | Yes | Yes | Yes |
| % Same-race ↑ uptake per new link | No | No | No | No | No | No | No | No | No |
| Women ↓ uptake per new link | Yes | Yes | Yes | Yes | Yes | Yes | Yes | Yes | Yes |
| Non-white ↓ uptake per new link | No | No | No | Yes | Yes | Yes | Yes | Yes | No |
| **Distal novelty** | | | | | | | | | |
| % Same-gender ↑ distality | Yes | Yes | Yes | Yes | Yes | Yes | Yes | Yes | Yes |
| Distality ↓ uptake per new link | Yes | Yes | Yes | Yes | Yes | Yes | Yes | Yes | Yes |
| **Novelty's relation with careers** | | | | | | | | | |
| Novelty ↑ faculty research | Yes | Yes | Yes | Yes | Yes | Yes | Yes | Yes | Yes |
| Novelty ↑ continued research | Yes | Yes | Yes | Yes | Yes | Yes | Yes | Yes | Yes |
| **Impactful novelty's relation with careers** | | | | | | | | | |
| Novelty ↑ faculty research | Yes | Yes | Yes | Yes | Yes | Yes | Yes | Yes | Yes |
| Novelty ↑ continued research | Yes | Yes | Yes | Yes | Yes | Yes | Yes | Yes | Yes |
| **Novelty discount on careers** | | | | | | | | | |
| Gender minorities novelty discount for faculty | No | Yes | Yes | Yes | Yes | Yes | No | Yes | Yes |
| Gender minorities novelty discount for cont. research | No | Yes | Yes | No | No | Yes | No | Yes | Yes |
| Racial minorities novelty discount for faculty | Yes | Yes | Yes | Yes | Yes | Yes | Yes | Yes | Yes |
| Racial minorities novelty discount for cont. research | Yes | Yes | Yes | Yes | Yes | Yes | Yes | Yes | Yes |
| **Impactful novelty discount on careers** | | | | | | | | | |
| Gender minorities impact discount for faculty | Yes | Yes | Yes | Yes | Yes | Yes | Yes | Yes | Yes |
| Gender minorities impact discount for cont. research | Yes | Yes | Yes | Yes | Yes | Yes | Yes | Yes | Yes |
| Racial minorities impact discount for faculty | Yes | Yes | Yes | Yes | Yes | Yes | Yes | Yes | Yes |
| Racial minorities impact discount for cont. research | No | No | No | No | No | No | No | No | No |

**Table S3.** *Novelty and impactful novelty correspond with publication productivity and impact. Descriptive analyses (linear regression models) where we regress total number of publications and accumulated citations (both logged) of students' work on novelty and impactful novelty. We use fixed effects for academic discipline, year of PhD graduation, and PhD university. We find that the novelty and impactful novelty positively relate to the number of publications and students' accumulated citations.*

|  | *log*(# Publications) | | | | | | *log*(# Citations) | | | | | |
|---|---|---|---|---|---|---|---|---|---|---|---|---|
|  | Coef. | S.E. | *p* | Coef. | S.E. | *p* | Coef. | S.E. | *p* | Coef. | S.E. | *p* |
| *log*(# New links) | 0.034 | 0.001 | 0.000 | | | | 0.030 | 0.002 | 0.000 | | | |
| *log*(Uptake per new link) | | | | 0.037 | 0.002 | 0.000 | | | | 0.054 | 0.002 | 0.000 |
| *log*(# Publications) | | | | | | | 1.337 | 0.002 | 0.000 | 1.333 | 0.002 | 0.000 |
| Observations | 532,077 | | | 425,318 | | | 464,920 | | | 374,415 | | |
| Inclusion criteria | Publishing | | | Publishing with nonzero novelty | | | Cited publication | | | Cited publication with nonzero novelty | | |

**Table S4.** *Concepts with their ten nearest, most-proximal neighbors in the embedding space. While there are no predefined or definitive tests for precisely quantifying what concept embeddings capture, we show here that concept embeddings capture semantic distances between concept quite effectively. To do this, we consider a few arbitrary concepts and look at their ten nearest neighbors in the embedding space as shown. By examining the set of nearest neighbors for these set of sample concepts, we note that the nearest neighbors are semantically similar to the focal concept. For instance, note that words which are similar to "syntax" include concepts like "grammar," "phrase structure," "semantics," and "word order" suggesting that concepts close to other concepts in this vector space captured through the embeddings effectively capture two concepts that are highly related substantively.*

| Concept | Ten nearest neighbors in the embedding space | | | | |
|---|---|---|---|---|---|
| gene | gene_encod | genes_involv | pathway_gen | gene_clust | gene_locus |
|  | genes_were_found | genes_loc | gene_rev | gene_set | regulatory_target |
| magnet | magnetic_field | local_magnet | magnetoresist | nonmagnet | ferromagnet |
|  | high_magnet | external_magnet | large_magnetic_field | weak_magnet | hard_axi |
| hiv | hiv_infect | hiv-infect | human_immunodeficiency_virus | hcv | hiv_transmiss |
|  | hiv-posit | gbv-c | haart | hiv_posit | hiv_diseas |
| fiscal | fiscal_polici | budgetari | revenues_and_expenditur | intergovernmental_gr | non-fisc |
|  | debt_servic | fiscal_stress | public_spend | local_fisc | state_fisc |
| christian | evangel | non-christian | theolog | catholic | christian_faith |
|  | judaism | nicen | jewish_peopl | american_protest | montanist |
| optic | qoct | wavelength-select | solid_immers | optical_filt | light_guid |
|  | coupled_cav | coherent_light | superlens | wavelength-tun | all-fib |
| topolog | binary_hypercub | vertex_and_edg | fat-tre | global_topolog | connection_matrix |
|  | sw-banyan | rectangular_du | physical_topolog | graph_properti | graph_represent |
| buddhist | buddhism | daoist | non-buddhist | taoist | theravada |
|  | tantric | pure_land | buddha | chinese_buddhist | neo-confucian |
| religion | religi | traditional_religion | american_civil_religion | manikkavacakar | christian |
|  | sikhism | salaf | relationship_between_religion | varkari | islamic_faith |
| oxygen | o2 | high_oxygen | molecular_oxygen | oxid | sulfur |
|  | low_oxygen | carbon_dioxid | presence_of_oxygen | amount_of_oxygen | peroxid |
| laser | q-switch | laser_puls | diode-pump | fs_puls | narrow_linewidth |
|  | diode_las | pump_sourc | 1064nm | laser_beam | femtosecond_las |
| electron | spin_degree_of_freedom | single-electron | spin_hall_effect | spin_accumul | charge_and_spin |
|  | conduction_electron | electron-volt | photoinject | coupled_quantum_wel | single_quantum_dot |
| proton | deuteron | methylene_proton | protonated_and_unproton | proton_transf | n-proton |
|  | d-channel | dominant_react | dehydron | hydrazyl | halogen_atom |
| photon | single_photon | polarization-entangl | entangled_photon_pair | nonlinear_cryst | photon_pair |
|  | spontaneous_parametr | single-photon | high_harmonic_gener | beamstrahlung | antibunch |
| oil | petroleum | asphaltene_cont | crude_oil | oil_extract | oil_sand |
|  | bitumen-deriv | linse | bitumen | liquefied_natural_ga | soybean_oil |
| tomb | sarcophagi | funerari | statu | monument | stela |
|  | palac | shrine | templ | necropoli | statuett |
| syntax | syntact | syntactic_structur | grammar | syntactic_and_semant | phrase_structur |
|  | grammat | syntactic_analysi | semant | grammatical_categori | word-ord |

**Supplementary References**